%% file: article.tex
\documentclass[11pt]{article}
\usepackage{xspace}
\usepackage{graphicx}
\usepackage{amsmath}
\usepackage{amssymb}
\usepackage{color}
\usepackage{subfig}

\input econfmacros.tex
\def\Title#1{\begin{center} {\Large {\bf #1} } \end{center}}

\begin{document}

\Title{TITUS: An Intermediate Distance Detector for the Hyper-Kamiokande Neutrino Beam}

\bigskip\bigskip


\begin{raggedright}  

{\it Pierre Lasorak \index{Lasorak, P} and Nick Prouse \index{Prouse, N.} for the TITUS working group,\\
PPRC, School of Physics and Astronomy\\
Queen Mary University of London\\
E1 4NS London, UK}\\

\end{raggedright}
\vspace{1.cm}

{\small
\begin{flushleft}
\emph{To appear in the proceedings of the Prospects in Neutrino Physics Conference, 15 -- 17 December, 2014, held at Queen Mary University of London, UK.}
\end{flushleft}
}

\section{Introduction}
TITUS (Tokai Intermediate Tank with Unoscillated Spectrum) is a proposal for an intermediate detector as part of the Hyper-Kamiokande~(HK) experiment~\cite{Abe:2011ts}. It will be located approximately 2~km from the J-PARC neutrino beam. TITUS is a cylindrical Cherenkov detector, filled with about 2~ktonne of gadolinium (Gd) doped water, aligned with \(2.5^{\circ}\) off-axis with respect to the neutrino beam. A magnetised iron muon detector is located at the downstream part of the tank to measure muons ranging out of the detector.
The Cherenkov effect allows detection and identification of electrons and muons produced in neutrino Charged Current (CC) interactions, and Gd allows the detection of possible outgoing neutrons. The primary goal of TITUS is to constrain the neutrino flux from the J-PARC beam that directly affects the sensitivity to CP violation at the far detector. The selection of the neutrino flux at the near detector is improved with respect to water Cherenkov-only tanks thanks to the Gd, that allows the capture of the final state neutrons from the neutrino-nucleon interaction and allows neutrinos and antineutrinos to be distinguished. A precise measurement of neutrino cross sections in water further helps the selection. TITUS can also be used for other physics purposes including detection of supernovae neutrinos, sterile neutrino studies and understanding the background for proton decay searches.

\section{Hyper-Kamiokande}
Hyper-Kamiokande is a proposed next generation neutrino oscillation experiment using a 1~megatonne water Cherenkov detector. It aims to study CP violation in the lepton sector by comparing the oscillation probabilities for neutrinos and antineutrinos~\cite{Abe:2015zbg}. To achieve this, the systematic uncertainties need to be greatly reduced from the current neutrino oscillation experiment at T2K. HK will also measure other neutrino oscillation parameters; it is expected to probe the proton life-time at an order of magnitude beyond the current limit~\cite{Abe:2011ts,Abe:2015zbg}, and it will be able to study astrophysical neutrinos.

\section{TITUS}
\begin{figure}[!ht]
\begin{center}
\subfloat[\(\nu_\mu\) flux for beam in neutrino mode]{
\includegraphics[width=0.33\textwidth]{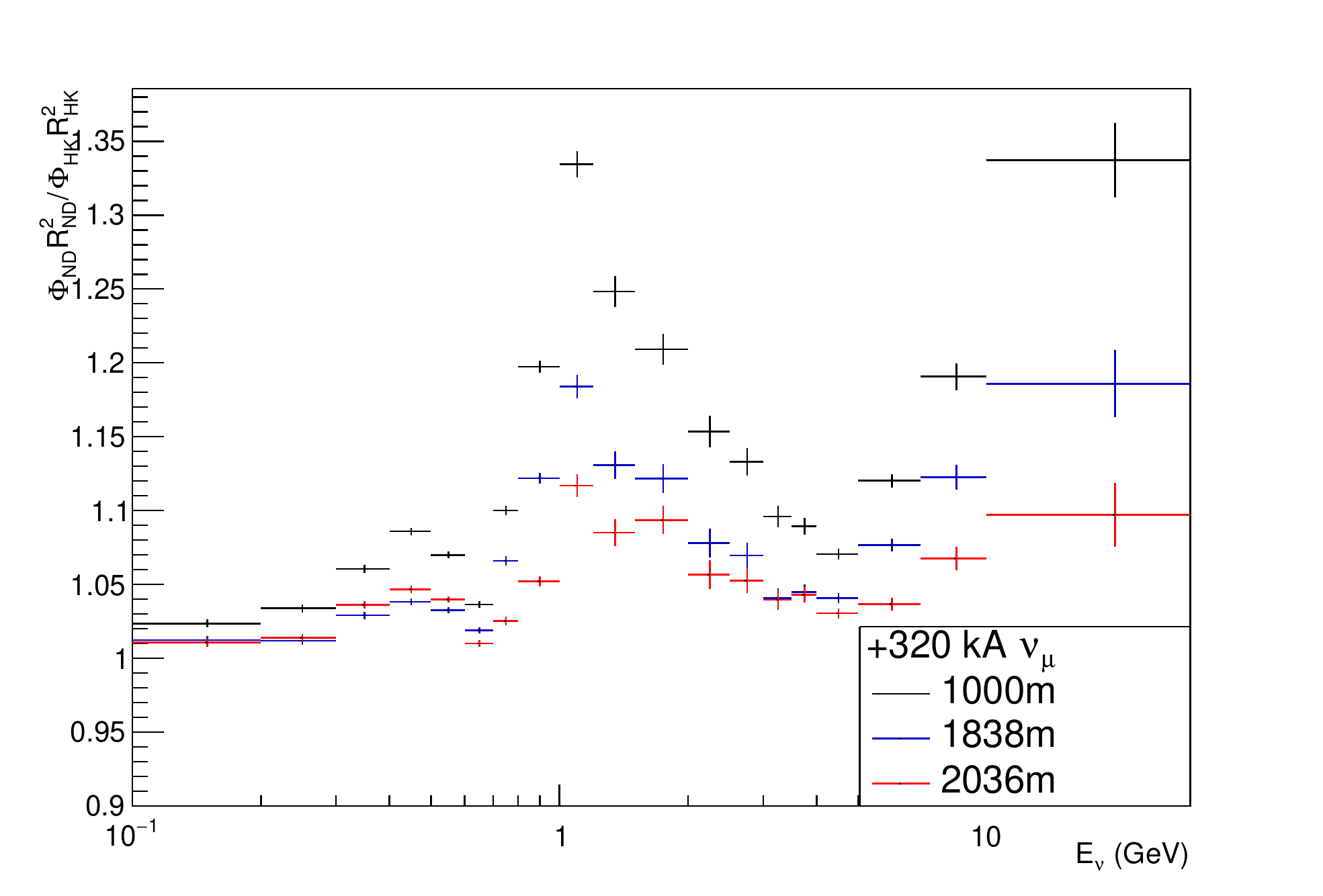}
}
\hspace*{0.05\textwidth}
\subfloat[\(\bar{\nu}_\mu\) flux for beam in antineutrino mode]{
\includegraphics[width=0.33\textwidth]{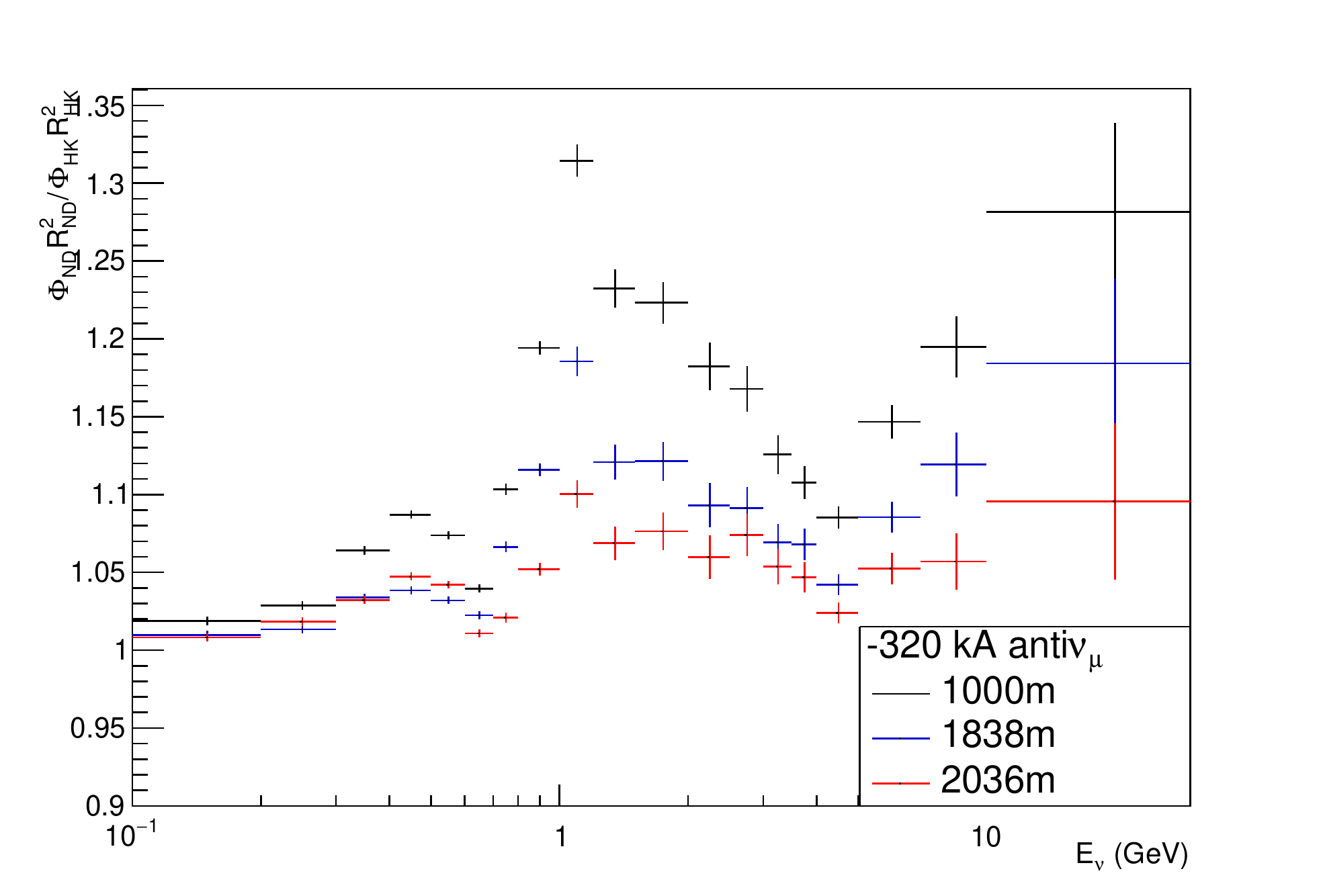}
}
\caption{Unoscillated flux ratios (Nominal HK / Near Detector) at baselines of 1000m, 1828m, and 2036m, for \(\nu_\mu\) with neutrino enhanced beam (left) and \(\bar{\nu}_\mu\) with antineutrino enhanced beam (right).}
\label{fig:fluxratios}
\end{center}
\end{figure}

To achieve the required reduction of the systematics uncertainties, precise measurements are required of both the unoscillated neutrino flux and the neutrino cross section. This can be achieved using TITUS, an intermediate detector between the J-PARC beam source and HK. By using the same target (water) and flux as HK, it is possible to cancel many of the differences between the near and far detectors. A smart selection using Gd helps to reduce the background. Figure~\ref{fig:fluxratios} shows the ratio of the flux for different baselines for TITUS and the flux at HK. The flux at \(\simeq 2\)~km is very similar to that at HK.
\subsection{Gadolinium doping}
\begin{figure}[!ht]
\begin{center}
\subfloat[Before neutron tagging]{
\includegraphics[width=0.28\textwidth]{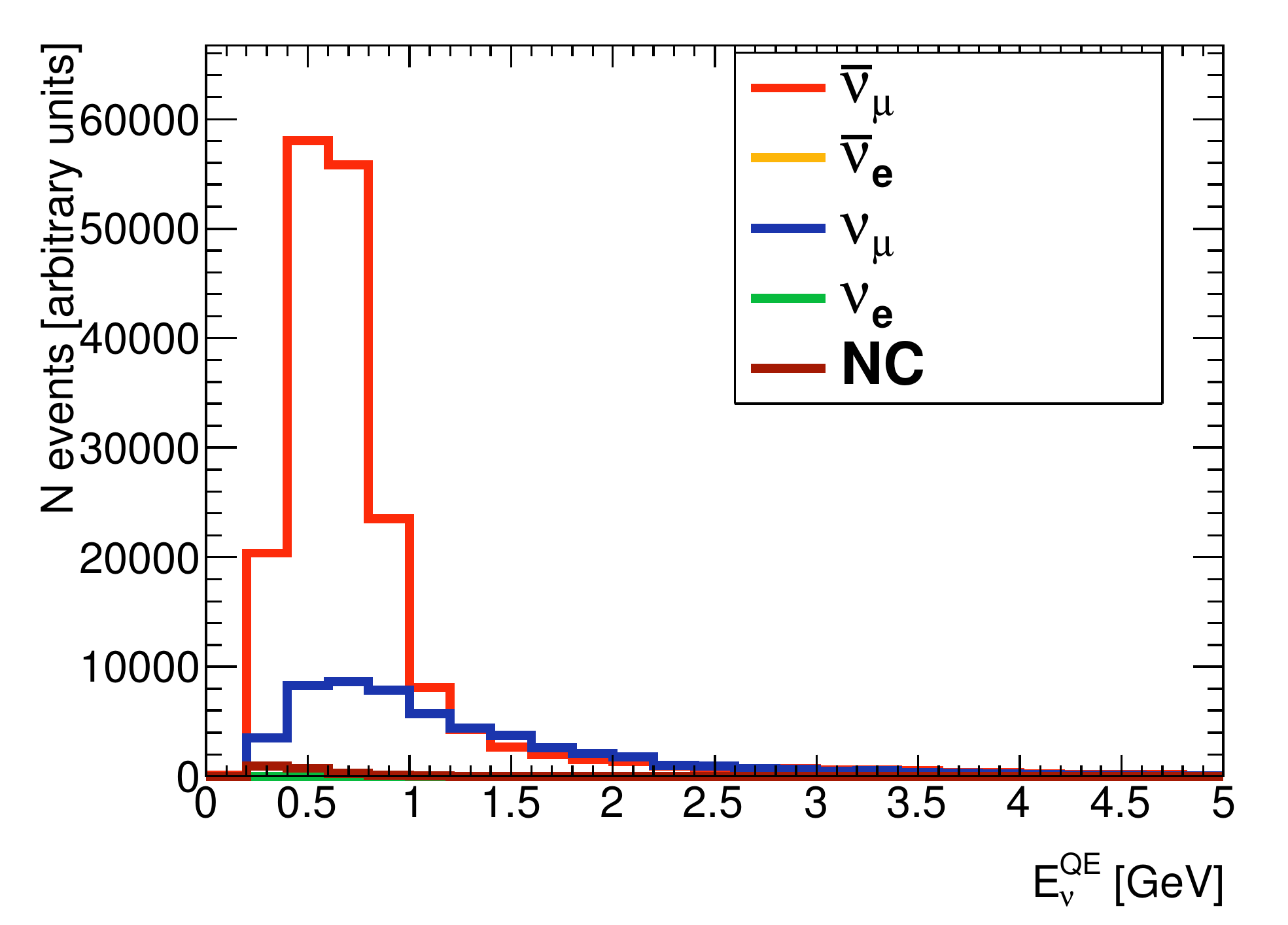}
}
\subfloat[No tagged neutron]{ 
\includegraphics[width=0.28\textwidth]{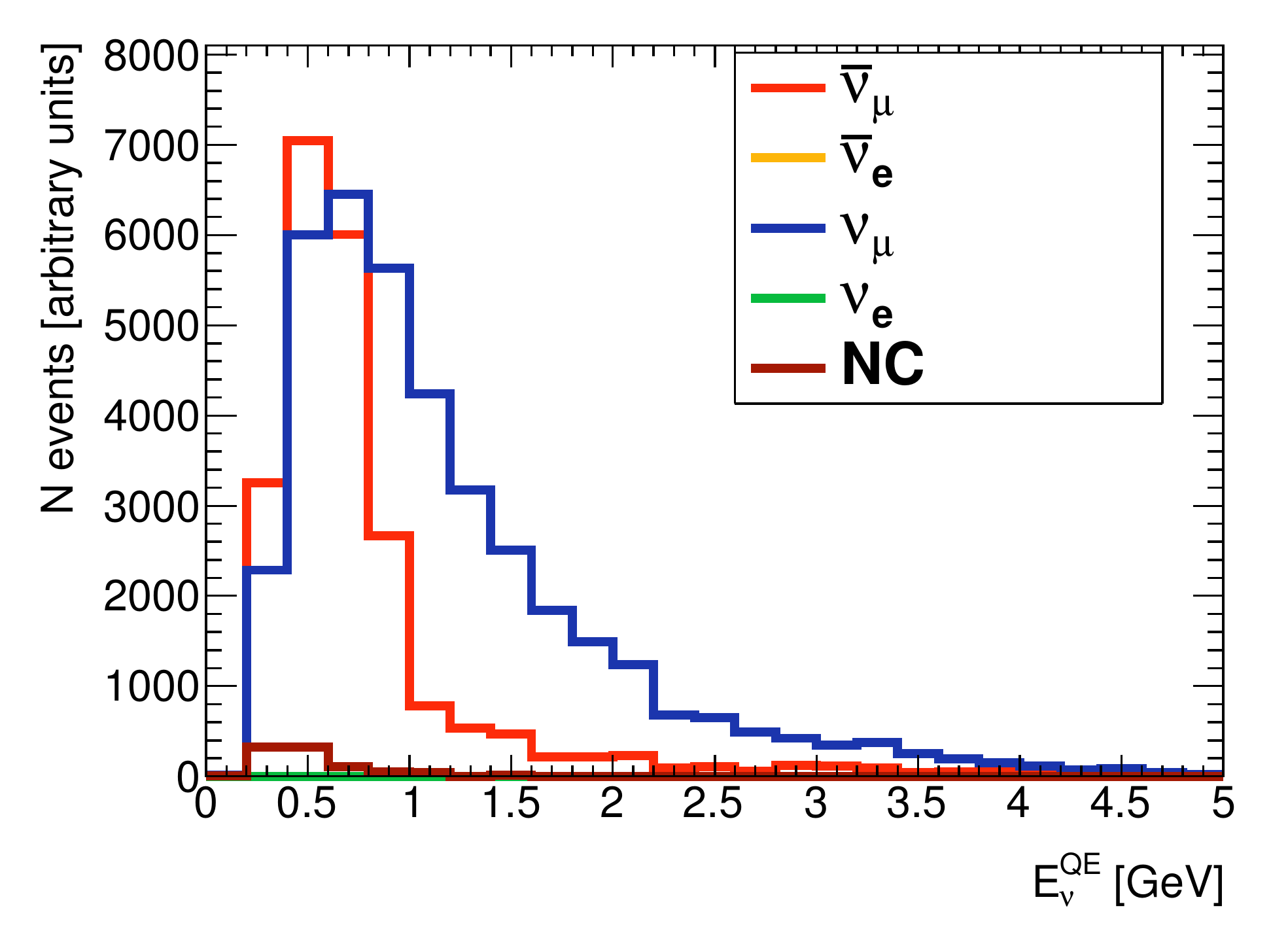}
}
\subfloat[Tagged neutron]{ 
\includegraphics[width=0.28\textwidth]{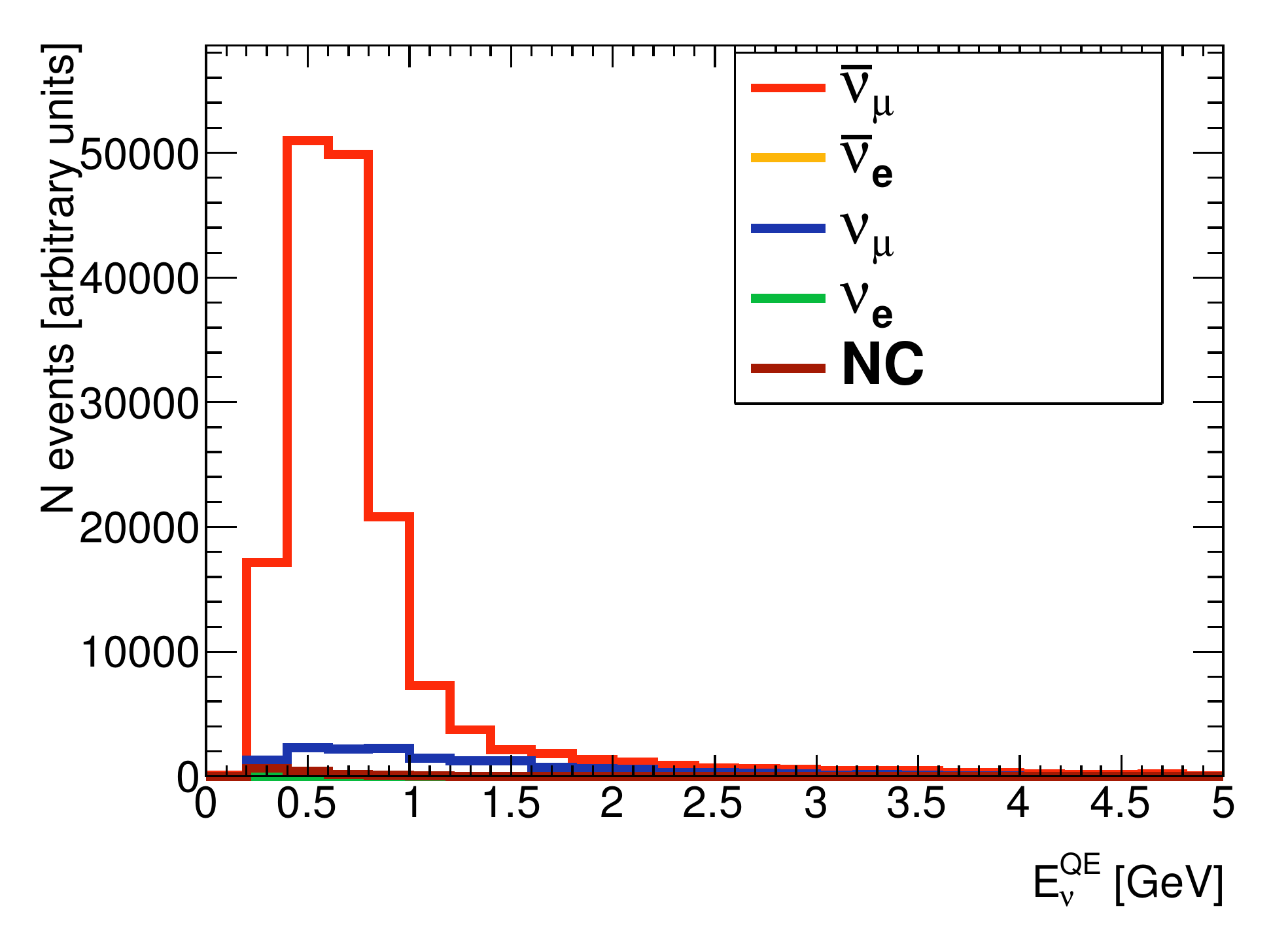}
}
\caption{The composition of the one muon-like ring sample in TITUS during anti-neutrino mode running. The effect of different neutron selections is shown. \label{fig:selections}}
\end{center}
\end{figure}

Doping the water with 0.1\% of Gd allows the detection of neutrons produced in neutrino interactions; this is realised because the neutron capture on Gd has a very high cross-section and produces a cascade of photons with total energy of about 8~MeV producing 4-5~MeV of visible energy that can be detected~\cite{Dazeley:2008xk}. For an oscillation analysis this can provide a very pure sample of CCQE interaction events both when a neutrino is interacting (producing no neutrons) or an anti-neutrino (producing 1 neutron). The effect on the spectrum of the selection can be seen in Figure~\ref{fig:selections}. R\&D is ongoing to monitor the feasibility and response of the detector when the water is doped with Gd~\cite{Renshaw:2012np, Anghel:2015xxt}.
\subsection{Photosensors}
Different types of photosensors are currently under investigation. Along with Photomultiplier Tubes (PMT), TITUS may include LAPPDs (Large Area Picosecond Photo Detectors), the next generation photosensors with improved timing resolution of the order of few tens of picoseconds and can reconstruct the hit position on the detector surface to within a few centimetres~\cite{Anghel:2013zxa}. Adding LAPPDs greatly improves the event reconstruction for low energy events (neutron capture on Gd). These detectors are currently being developed.

\subsection{Magnetised Muon Range Detector}
\begin{figure}[!ht]
\centering
\begin{minipage}{.45\textwidth}
\centering
\includegraphics[width=0.8\columnwidth]{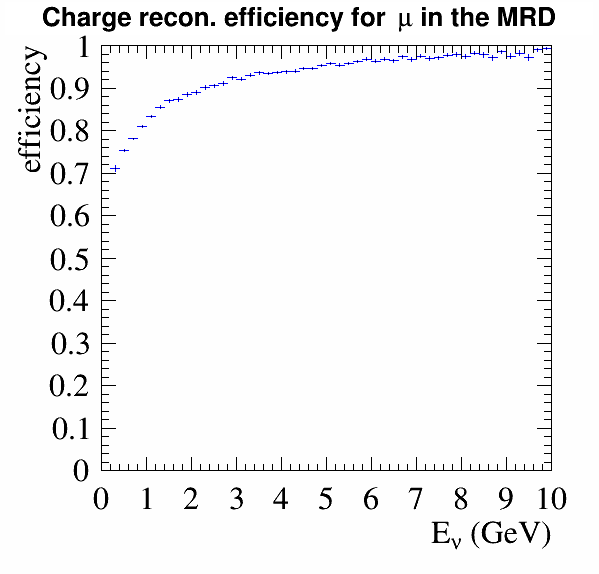}
\captionof{figure}{MMRD charge reconstruction efficiency  for muons coming from the interaction of the neutrinos with the tank.}
\label{fig:MMRD}
\end{minipage}
\hspace*{0.05\textwidth}
\begin{minipage}{.45\textwidth}
\centering
\includegraphics[width=0.9\columnwidth]{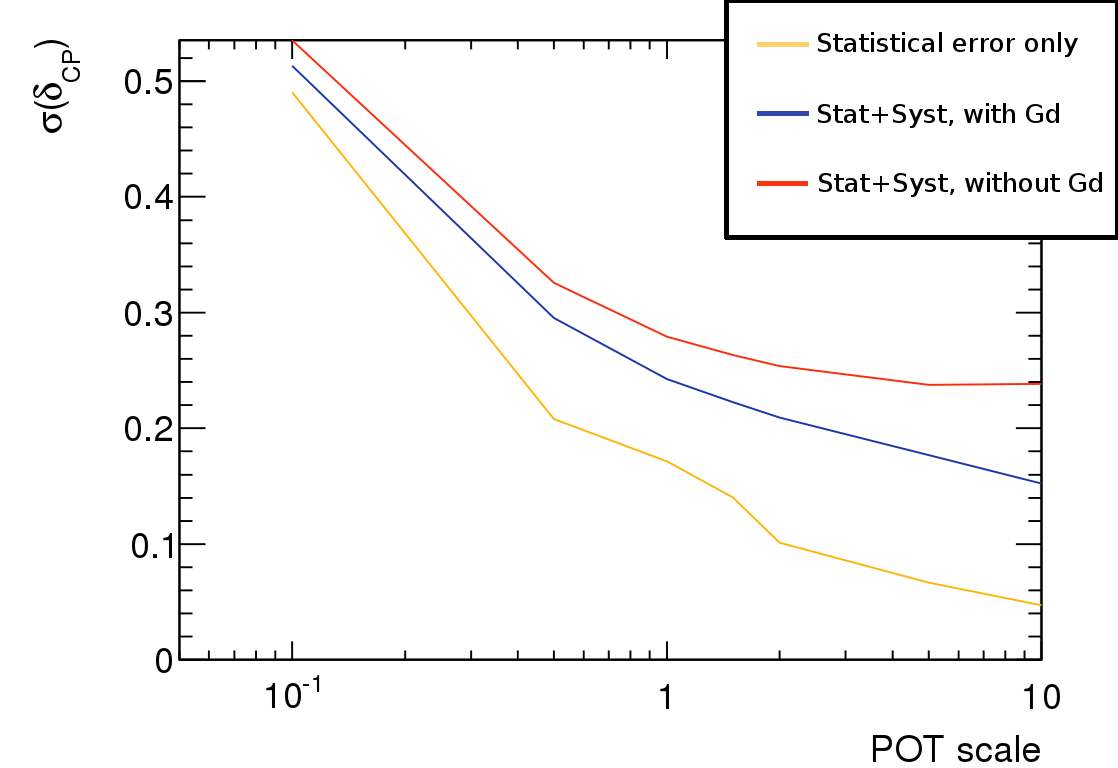}
\captionof{figure}{Scaling of the error on \(\delta_{\text{CP}}\), where a POT scale of one is after 10~years of operation with a beam of 750~kW and equal splitting between neutrino and anti-neutrino mode beam. The study was realised assuming \(\sin^2(2\theta_{13})=0.095\) and \(\delta_{\text{CP}}=0\).}
\label{fig:CPsensitivity}
\end{minipage}
\end{figure}

Due to the size of TITUS, about 18\% of the muons coming from beam neutrino interactions escape the tank. These muons come from neutrinos in the higher end of the spectrum. It is therefore important to quantify their energy after they ranged out of the detector to help in understanding the high energy background.
A Magnetised Muon Range Detector (MMRD) with magnetic field of 1.5~T can provide energy and charge reconstruction.
Figure~\ref{fig:MMRD} shows the charge reconstruction efficiency dependent on neutrino energy. 
Combined with the neutron tagging this could give very high purity samples as well as providing a method for validating and calibrating the neutron tagging.
\section{CP violation sensitivity}
Due to the very high discrimination of TITUS described in the previous section, the sensitivity of HK to CP violation is increased. The addition to the CP violation fit at HK of the intermediate detector sample, and in particular with the neutron-tagged sample, where detector systematics including neutrino cross-section and nucleon final state interaction uncertainties have been included, leads to a significantly decreased time to discovery by a reduction in the total systematic error at the far detector, as seen in Figure~\ref{fig:CPsensitivity}.




\section{Summary}
The addition of TITUS, a 2~ktonne Gd-doped water Cherenkov detector with magnetised muon range detector, located approximately at 2~km from J-PARC, to the HK project will allow precise measurements of the unoscillated spectrum of the J-PARC neutrino beam. The 0.1\% doping of Gd allows for a detectable signal from neutron capture, and thus the discrimination of neutrino and antineutrino interactions as well as inputs into neutrino cross-section measurements. The downstream MMRD provides a second method for neutrino/antineutrino discrimination through charge reconstruction of muons exiting the tank and also provides energy reconstruction for these muons. These features allow TITUS to significantly reduce systematic errors in CP violation measurements at HK, providing increased sensitivity and reduced time to discovery.



\end{document}

%% file: econfmacros.tex
\definecolor{Red}{rgb}{1,0,0}
\definecolor{Green}{rgb}{0,1,0}
\definecolor{Blue}{rgb}{0,0,1}
\definecolor{Black}{rgb}{0,0,0}

\def\beq{\begin{equation}}
\def\eeq#1{\label{#1}\end{equation}}
\def\eeqn{\end{equation}}

\def\beqa{\begin{eqnarray}}
\def\eeqa#1{\label{#1}\end{eqnarray}}
\def\eeqan{\end{eqnarray}}

\let\bar=\overbar

\def\Dslash{\not{\hbox{\kern-4pt $D$}}}
\def\dslash{\not{\hbox{\kern-2pt $\del$}}}

\def\msb{{\bar{\ssstyle M \kern -1pt S}}}